\documentclass[aip, apl, amsmath, amssymb, floatfix, preprint, twocolumn]{revtex4-1}

\usepackage{graphicx}
\usepackage{dcolumn}
\usepackage{bm}
\usepackage[utf8]{inputenc}
\usepackage[T1]{fontenc}
\usepackage{mathptmx}
\usepackage{etoolbox}
\makeatletter
\def\@email#1#2{%
 \endgroup
 \patchcmd{\titleblock@produce}
  {\frontmatter@RRAPformat}
  {\frontmatter@RRAPformat{\produce@RRAP{*#1\href{mailto:#2}{#2}}}\frontmatter@RRAPformat}
  {}{}
}

\makeatother

\begin{document}


\title[Enhancing SNSPD performance via a Silicon passivation layer]{Enhanced Superconducting Nanowire Single Photon Detector Performances using Silicon capping}
\author{C. Klein }\altaffiliation{a}
\author{S. Cohen }\altaffiliation{a}
\affiliation{KTH royal institute of technology,  Quantum-Nano-Photonics,  Stockholm, 10450, Sweden}
\author{T. Descamps}
\affiliation{KTH royal institute of technology,  Quantum-Nano-Photonics,  Stockholm, 10450, Sweden}
\author{A. Iovan}
\affiliation{KTH royal institute of technology, Quantum-Nano-Photonics, Stockholm, 10450, Sweden}
\author{P. Zolotov}
\affiliation{Single Quantum BV, Rotterdamseweg 394,  2629HH Delft, Netherlands}
\author{P. Vennéguès}
\affiliation{Université Côte d'Azur, CNRS, CRHEA, Sophia Antipolis, France}
\author{I. Florea}
\affiliation{Université Côte d'Azur, CNRS, CRHEA, Sophia Antipolis, France}
\author{F. Semond}
\affiliation{Université Côte d'Azur, CNRS, CRHEA, Sophia Antipolis, France}
\author{*V. Zwiller}
\email{zwiller@kth.se.\\ \ a  $^{a)}$ Equal contribution : S.C and C.K. contributed equally to this work}
\affiliation{KTH royal institute of technology,  Quantum-Nano-Photonics,  Stockholm,  10450,  Sweden}

\date{\today}

\begin{abstract}
Niobium Titanium nitride (NbTiN) based superconducting nanowire single photon detectors (SNSPDs) are known for their high performance across a wide spectral range, from the X-ray to the mid-infrared. Nonetheless, fabrication challenges and performance degradation attributable to surface oxidation and lack of uniformity in films thinner than 5 nm remain a significant barrier for achieving high-quality detectors. In this work, we study the influence of a Silicon capping layer on film properties and on the performance of SNSPDs. A Silicon capping layer effectively suppresses oxidation and increases the superconducting transition temperature. This enables superconductivity in films as thin as 3 nm at 3 K, increases critical current in patterned nanowires and significantly extends the saturation plateau from the visible to the near infrared (up to 2050 nm): These detectors maintain sub-50 ps timing jitter, even for nanowires as wide as 250 nm and with detection areas of 20x20$\mu m^{2}$. 
Our results establish that thinner films protected by a capping layer allow for the fabrication of wider wires, decreasing nanofabrication challenges and extending the operating temperature range for efficient single photon detection. 

\pagebreak{}
\end{abstract}
\maketitle 

In recent years, SNSPDs emerged as a leading platform for high-performance photon detection, driven by their remarkable system detection efficiency (SDE)\cite{Chang2021}, ultra-low dark count rates (DCR)\cite{Shibata2017}, low timing jitter\cite{Korzh2020}, and high maximal count rates\cite{Craiciu2023}. These performance attributes have positioned SNSPDs as a key enabling technology across a range of applications, including Light Detection And Ranging (LiDAR)\cite{Staffas2024}, photoluminescence spectroscopy \cite{Morozov2021}, optical microscopy\cite{Tamimi2024}, and quantum communication\cite{Wengerowsky2020}. Conventional SNSPDs are typically fabricated from thin superconducting films, most commonly Niobium Nitride (NbN), Niobium Titanium Nitride (NbTiN), Tungsten Silicide (WSi) or Molybdenum Silicide (MoSi), which are patterned into narrow, meandered nanowires,  typically <100 nm in width and <10 nm in thickness. This nanoscale architecture comes with several limitations. Due to the small feature size, we face challenges related to yield and fabrication imperfections. Moreover, we typically rely on electron beam lithography, a technique that, while precise, is time-consuming and costly. Scaling SNSPDs to cover larger detection areas requires extending the length of the nanowire significantly, which, due to the high kinetic inductance of NbTiN, leads to increased timing jitter and prolonged recovery times: factors that limit their temporal performance\cite{Calandri2016}. To mitigate these issues, recent efforts focused on fabricating SNSPDs with nanowire widths up to 20 $\mu m$\cite{Yabuno2023}, \cite{Zolotov2023} as they offer several fabrication- and performance-related benefits\cite{Xu2021}.  Wider wires are easier to fabricate, less sensitive to lithographic imperfections,  and exhibit reduced kinetic inductance, resulting in shorter recovery times\cite{gol2001picosecond}.  However,  increasing the wire width typically requires a reduction in film thickness to maintain high enough resistance levels and thus saturated detection efficiency\cite{Smirnov2018}.  For NbTiN,  this motivates the use of films with thicknesses approaching 3 nm. 
However, decreasing film thickness reduces the transition temperature and makes superconducting properties highly sensitive to film inhomogeneities causing degraded device performance.\cite{Smirnov2018}   As reported in \cite{Zhang2018}, upon atmospherical contact, NbTiN develops a native oxide layer within hours to a thickness of the order of 1 nm. Therefore, in-situ passivation is crucial for minimizing oxidation,
particularly detrimental in films thinner than 5 nm,  where even a few atomic layers of oxide can significantly degrade superconducting properties\cite{Zhang2018}. 
In this work,  we systematically investigate the effect of a Silicon capping layer on the superconducting and optical response of  NbTiN films as well as the performance of SNSPDs patterned from these films.  Silicon was selected due to its transparency in the near-infrared spectral range \cite{Dorenbos2008}, low cost\cite{Feng2022}, compatibility with standard semiconductor processing\cite{Wang2018}, and its known function as an oxygen diffusion barrier\cite{Murarka2005}. We deposited films with NbTiN thicknesses ranging from 3 nm to 9 nm and characterized the critical temperature as a function of thickness and compared films with and without a Si capping layer.  We then fabricated SNSPDs from these films and evaluated their performances across visible and near infrared wavelengths.

Similar to Ref\cite{Zichi2019}, superconducting NbTiN films were sputtered on a SiO$_{2}$ (270 nm) / Si (500 µm) wafer.  Precise control over film stoichiometry was achieved by adjusting the sputtering powers applied to the Nb and Ti targets. Film thicknesses were monitored in situ using a calibrated quartz crystal. To mitigate surface oxidation, a 5 nm Silicon capping layer was deposited in situ after NbTiN growth.  Structural and morphological characteristics of the films were investigated using a combination of advanced electron and scanning probe microscopy techniques. High-resolution scanning transmission electron microscopy (STEM), coupled with energy-dispersive X-ray spectroscopy (EDX), was employed to assess the microstructure and map the spatial distribution of constituent elements with nanoscale resolution. Complementary surface characterization was performed using atomic force microscopy (AFM) to quantify the surface roughness, and scanning electron microscopy (SEM) to evaluate the surface morphology and film continuity. 
To understand the impact of the Si capping layer on the structural properties of NbTiN,  scanning electron microscopy (Fig 1.a),  Atomic force microscopy (Fig 1.b), and scanning transmission electron microscopy (Fig 1.c) in conjunction with energy-dispersive X-ray spectroscopy (Fig 1.d) were conducted. 
\begin{figure*}[!htbp]
    \includegraphics[scale=1]{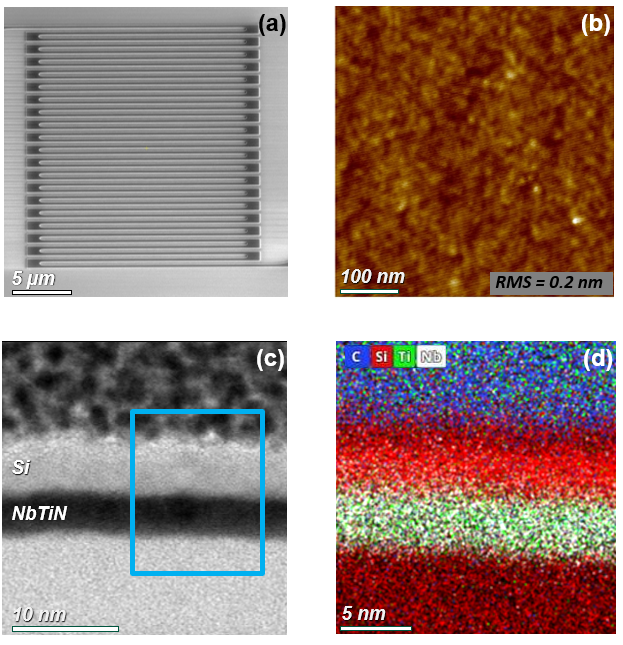}%
    \caption{\label{fig1}(a)  SEM image of a meander-shaped superconducting nanowire detector made from a 4 nm thick NbTiN film capped with 5 nm Silicon layer  (b) AFM image of a 3 nm  NbTiN thin film capped with 5nm Si, showing the surface morphology and extracted roughness of the NbTiN layer. (c)  BF-STEM image of a 4 nm NbTiN film capped with a 5 nm Silicon layer,  displaying the layer interfaces and showing the multilayer stack. (d) EDX elemental maps acquired from the region shown in Fig 1(c), confirming the spatial distribution and chemical separation of Nb (white), Ti (green), and Si(red).}
\end{figure*}
Fig 1. (a) presents a SEM image highlighting the nanowire's width uniformity and minimal edge roughness over large device areas (20 × 20 µm²). Fig 1.(b), shows the surface morphology of a 3 nm NbTiN film. From this we extracted a roughness of 0.2 nm (RMS).  Fig. 1(c) shows a  Bright-Field (BF) STEM image used to verify that a 5 nm silicon layer is deposited on top of the NbTiN/SiO$_2$/Si stack. As seen in Fig. 1(c), the NbTiN films grown at room temperature exhibit a polycrystalline structure, consistent with previously reported characteristics of room-temperature deposited NbTiN\cite{zhang2024effect}.   
Given that the Si layer was deposited via sputtering at room temperature, it is most likely amorphous\cite{zahid2021growth}. Furthermore, the Si capping layer has likely undergone partial oxidation to form native SiO$_2$, since even brief air exposure can initiate surface oxidation, a behavior also observed in Si capping layers on thin films used for oxidation protection \cite{van2013difference}. Film thicknesses were independently verified via X-Ray Reflectivity (XRR), with data fitted using the GenX software to ensure cross-method consistency. Fig. 1 (d) shows EDX elemental maps, confirming the distinct spatial separation of Silicon (red) from Nb and Ti (white and green, respectively), which clearly validates the presence of three well-defined layers. Hence, STEM analyses confirm the presence of a Silicon capping layer atop the NbTiN film. AFM and SEM characterizations emphasize the high surface quality, and fabrication consistency of the capped devices.

\begin{figure*}[!htbp]
    \includegraphics[scale=0.9]{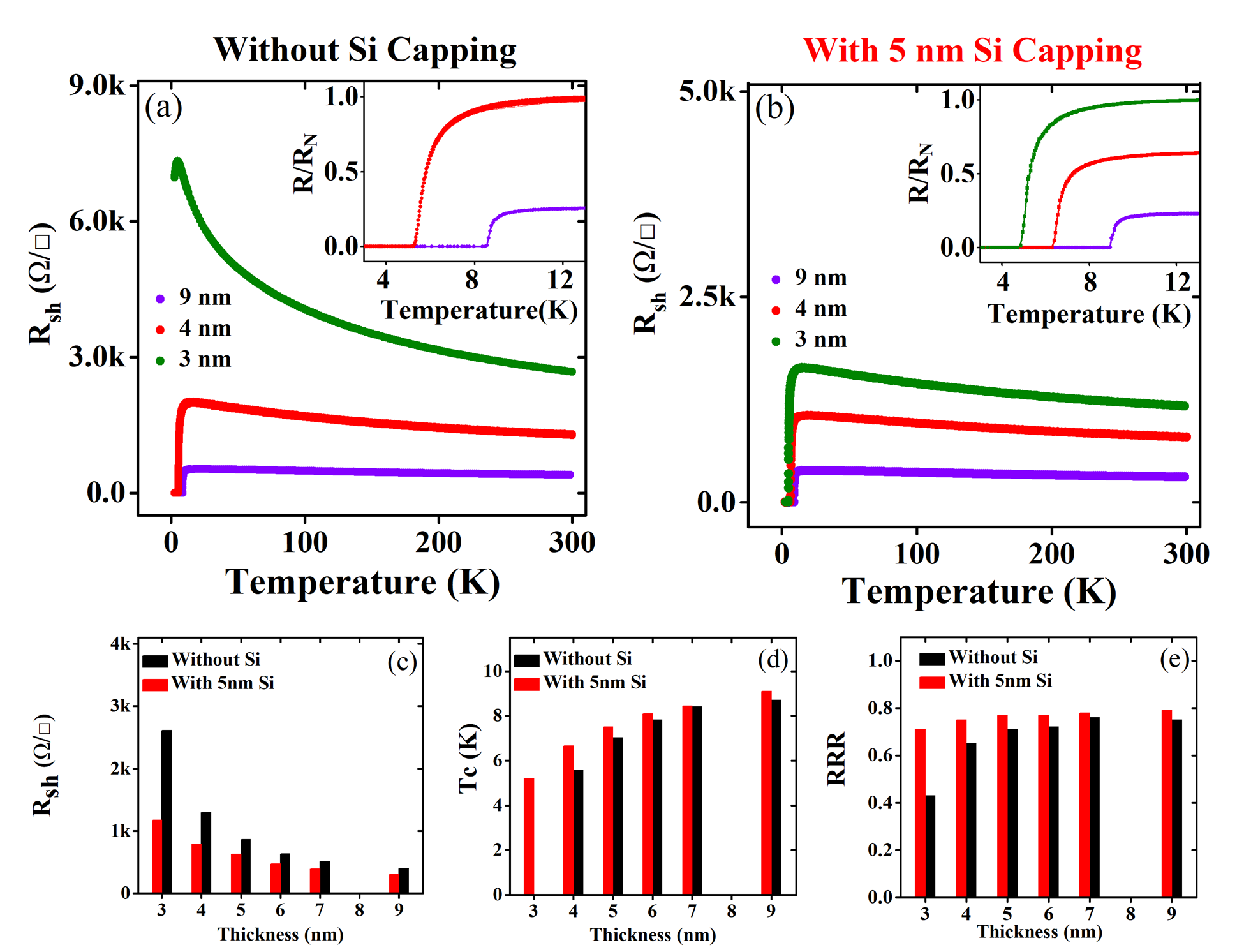}
    \caption{\label{fig2}(a) Critical temperature $T_{c}(K)$ versus $R_{sh} (\Omega/\Box)$ as a function of thickness (3–9 nm) for samples without Si capping. The inset shows data for superconducting samples (4–9 nm) within a narrower temperature range near their superconducting transitions. (b) Critical temperature $T_{C}(K)$ Versus $R_{sh} (\Omega/\Box)$ as a function of  thickness (3-9 nm) for samples with a 5 nm Si capping layer. The inset shows the superconducting samples (3–9 nm) within a narrower temperature range near their superconducting transitions. (c) Comparison of $R_{sh}$ ($\Omega/\Box$) as a function of NbTiN layer thickness (3–9 nm) without (Black) and with 5 nm Si capping (red). (d) Comparison of $T_{c}(K)$ as a function of film thickness (3–9 nm) without (Black) and with 5 nm Si capping layer (red). (e) RRR as a function of thickness (3–9 nm) without (Black) and with Si capping (red).} 
\end{figure*}

As mentioned previously, increasing the wire width requires reducing the film thickness to maintain high detection efficiency. Accordingly, NbTiN superconducting films were deposited with thicknesses ranging from 3 to 9 nm, as shown in Fig 2. (For the full measurement, see supplementary (Fig.1 )). For each film, two critical parameters: sheet resistance $R_{sh}$ and transition temperature $T_{C}$ were measured. For electrical characterization, the films were diced into square specimens and mounted in a four-point probe configuration. The sample temperature was swept down to 3 K in a closed cycle cryostat. The primary electrical parameters assessed included the room temperature sheet resistance ($R_{sh}$), the critical temperature ($T_{C}$), and the residual resistance ratio (RRR). 
The critical temperature ($T_{C}$) was measured at the temperature where the sheet resistance reaches 50\% of its value at 20 K. We consider the ratio of the sheet resistance measured at 300 K to that at 20 K as the RRR\cite{Smirnov2018}. 
As for the fabrication of the SNSPDs, we used a 50 keV electron-beam lithography tool to pattern MaN2401 resist from Micro Resist, followed by reactive ion etching with a gas mixture containing $SF_{6}$ and $O_{2}$, to define the nanowire structures. The meanders covered an area of 20×20 $\mu m^{2}$ with a fill factor of 50 \% and line widths of 100 nm. Device characterization was conducted at 3.2 K in a closed-cycle cryostat, where the detectors were flood-illuminated at wavelengths of 650 nm, 1310 nm,  and 1550 nm. The relationship between $R_{sh}$ and  $T_{C}$ is presented in Fig. 2(a). Two clear trends are observed: First, $T_{C}$ decreases as the film thickness decreases; second,  $R_{sh}$ increases with decreasing thickness. Notably, the 3 nm NbTiN film exhibits no superconducting transition above 3 K. 
While increasing $R_{sh}$ is beneficial for extending the saturation plateau in SNSPDs\cite{Zolotov2023},  this comes at the cost of reduced $T_{C}$, particularly for films thinner than 4 nm. At these thicknesses, especially in the order of magnitude of the coherence length of NbTiN, the superconducting properties may vary due to disorder and oxidation. To study the effect of the non-superconducting oxide layer on electrical parameters, NbTiN thin films were sputtered with thicknesses ranging from 3 to 9 nm. In a separate batch, to prevent oxidation, films were sputtered using the same protocol and also capped in situ with a 5 nm thick Silicon layer. 
Although various Si-capping layer thicknesses (3, 5, and 10 nm) were tested, no significant differences were observed in $T_{C}$ and $R_{sh}$ between thicknesses of 3nm and 5nm, supporting our hypothesis that the primary effect arises from the inhibition of oxidation in the NbTiN layer. From this measurement we also conclude that the difference in behaviour is not attributable to a change in surface strain. Additional measurements across different Si capping layer thicknesses can be found in the supplementary materials (Fig 2.). As a result, we selected 5 nm as a representative Si thickness for subsequent experiments. This compromise allows etching the film using the same optimized procedure as uncapped films while having thick enough capping to maintain superconductivity.
Fig 2.(b) shows the relationship between $R_{sh}$ and $T_{C}$ with capped films (5 nm Si). $T_{C}$ decreases as the NbTiN film thickness decreases. In contrast to uncapped samples, superconductivity was observed in films thinner than 4 nm at 3.2 K, highlighting the effectiveness of Silicon passivation in preserving superconductivity. A comparison of $R_{sh}$ as a function of film thickness for all films (3-9 nm) with (red) and without (black) Silicon is presented in Fig. 2c.  Without Silicon capping (black), $R_{sh}$ increases as the thickness decreases. However, after the Silicon capping layer (red), $R_{sh}$ is lower across all thicknesses, with the most significant improvement observed for films thinner than 5 nm. For instance, at a NbTiN film thickness of 3 nm, the sheet resistance ($R_{sh}$) decreased from $\sim$2500 to  $\sim$1500 $(\Omega/\Box)$,  which we attribute to the improved quality of the film due to the Silicon capping.  
Fig 2.(d) shows a comparison of $T_{C}$ as a function of thickness. Fig 2.(e) supports improved film quality through Si capping as RRR improves significantly across all thicknesses (3-9 nm)\cite{Krishnan2011}. 

\begin{figure*}[!htbp]
    \includegraphics[scale=0.74]{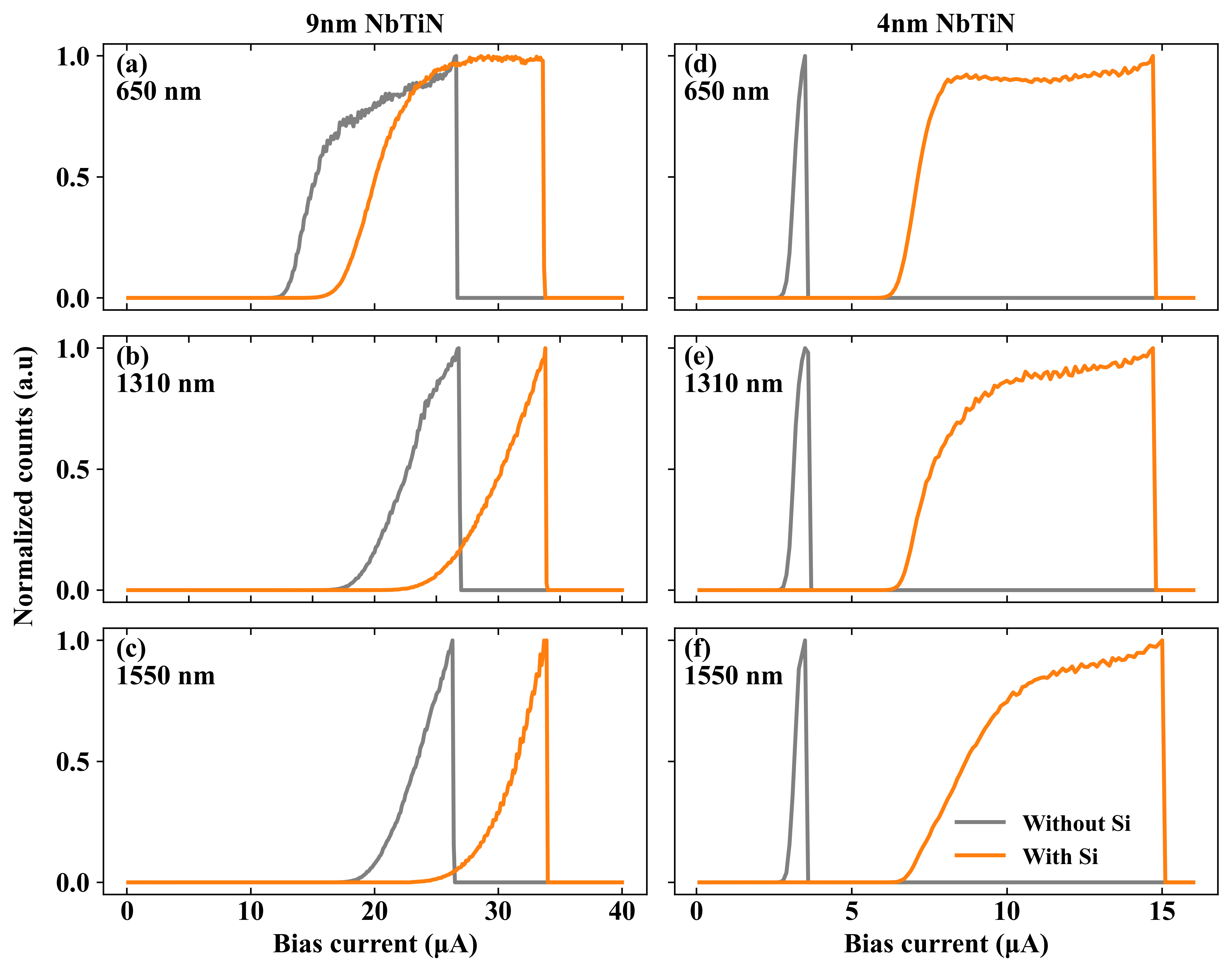}%
    \caption{\label{fig3}(a-c) Normalized counts rate of SNSPDs fabricated from 9 nm NbTiN (a-c) and 4 nm NbTiN (d-f) with a 5nm Silicon capping layer (orange) and without a capping layer (grey).  All nanowires were patterned with a 100 nm wire width, a fill factor of 50\% and a detector size of 20x20$\mu m^{2}$. Three more sets of devices fabricated in the same fashion exhibiting similar behavior can be found in supplementary (Fig.3) .}%
\end{figure*}

To understand the impact of Silicon capping on the performance of SNSPDs, we fabricated two sets of devices using NbTiN films with respective thicknesses of 9 nm and 4 nm,  each with and without a 5 nm Silicon capping layer. Fig.  3(a–c) shows the normalized count rate saturation curves for 100 nm-wide nanowires at wavelengths of 650 nm,  1310 nm,  and 1550 nm. For the 9 nm-thick devices, both uncapped (grey curves) and Silicon-capped (orange curves) detectors exhibit saturation at 650 nm (Fig.3a). However, at 1310 nm and 1550 nm (Figs. 3.b,c), both uncapped and capped devices show a significant reduction in the normalized count rate, failing to reach saturation. In all cases, the Silicon-capped devices display a higher critical current. A second set of detectors was fabricated using 4 nm-thick NbTiN, again with and without the Silicon capping layer,  Figs.3(d–f) present the corresponding data. Across all tested wavelengths, the uncapped devices exhibit poor saturation behavior, whereas the 4nm Si-capped devices demonstrate well-defined saturation even at longer wavelengths and shows higher critical currents. These results suggest that the Silicon capping layer serves as an effective surface passivation barrier, mitigating oxidation-induced degradation and thereby preserving superconducting properties. 
\begin{figure*}[!htbp]
    \includegraphics[scale=0.75]{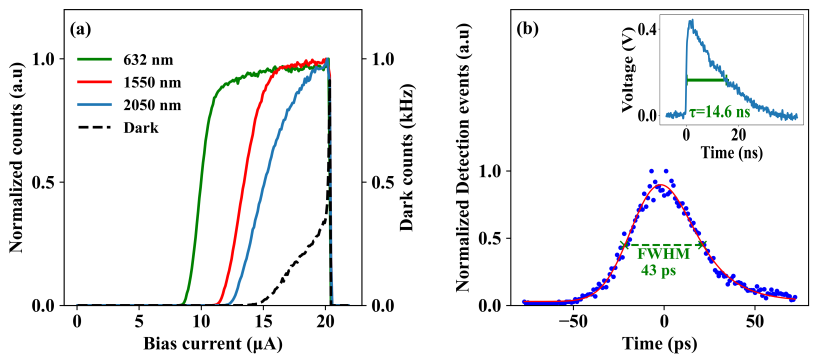}%
    \caption{\label{fig4}(a) Normalized count rate of SNSPDs fabricated from 4 nm-thick NbTiN films with 250 nm nanowire width and 5 nm Si capping layer, measured at 178 mK. (b) Timing jitter (main panel) measured at 1550nm, using an APE picoEmerald pulsed laser. The numerical value of the timing jitter was measured by fitting an exGaussian curve (red curve) to the time-correlated single-photon counting histogram (blue curve), resulting in a full width at half maximum (FWHM) jitter of 43 ps. Inset of Fig.4(b) displays a detection pulse with a reset time constant of 14.6 ns}%
\end{figure*}

To explore the advantages of Si-capping for the detection of photons at longer wavelengths as well as  for wider nanowires, we fabricated devices from 4 nm NbTiN films capped with 5 nm Si. These films were patterned into 250 nm wide nanowires covering an area of 20x20$\mu m^{2}$. To probe the SNSPDs at a wavelength of 2050 nm we loaded the samples into a different cryostat, also under flood illumination, explaining the different measurement temperature of 178mK. Saturation curves for the 250nm wide nanowires measured in the 3.2K cryostat can be found in the supplementary material (Fig.4). Fig.4(a) presents normalized count rates of a  device measured at 178 mK for wavelengths of 632 nm,  1550 nm,  and 2050 nm.  Across all wavelengths,  the detector demonstrate saturated detection efficiency. Notably, this performance extends into the near infrared regime for a detector covering an active area approximately five times larger (20x20$\mu m^{2}$)  than commercially available SMF-coupled detectors. 
The current state of the art\cite{Yabuno2025} shows efficient photon detection at 2000 nm with 20 $\mu$m wide strips. Such devices might also benefit  from the capping method presented here.
Beyond detection efficiency, the temporal resolution of the SNSPD remains a critical performance metric. Ref.\cite{Chang2019} reported sub-20 ps jitter on a similarly sized detector at 1064 nm using cryogenic amplification. We evaluated the timing jitter using a picosecond-pulsed laser at 1550 nm and a 4.5 GHz bandwidth oscilloscope. As shown in Fig. 4(b), the timing jitter of the SNSPD was determined by fitting an exponentially modified Gaussian curve to the time-correlated single-photon counting histogram, yielding a full width at half maximum (FWHM) jitter of 43 ps. Since the detector covers an area of 20x20$\mu m^{2}$, operates without cryogenic amplification, using lossy wiring (9 dB attenuation at 1 GHz), and that the jitter is measured at 1550 nm, it  shows that the Silicon capped detectors can be used for applications where timing jitter is important. The inset of Fig. 4(b) shows a detection pulse with reset time constant of 14.6 ns, a value to be expected for such a long detector. 
This demonstrates strong performance for wider nanowire geometries with films < 5 nm. 
\section*{Conclusion}
We studied the influence of Si capping on the performance of SNSPDs fabricated from thin NbTiN films.  We show that Silicon capping increases Tc, prevents surface oxidation,  preserves superconductivity at 3 K for films as thin as 3 nm, and improves film quality.  Importantly, Silicon capping increases the critical current and enables high detection efficiency in SNSPDs at longer wavelengths for wider and longer nanowire geometries,  while maintaining excellent timing resolution with sub-50 ps jitter.  These results establish Silicon capping not merely as a protective barrier,  but as a crucial enabler for extending the operational range and relaxing the fabrication requirements for SNSPDs. 
 
\bibliography{Capping_paper/bibliography/references}

\end{document}